\documentclass[structabstract, bibyear]{aa}  
\usepackage{graphicx}
\usepackage{txfonts}
\usepackage{color}

\begin{document}

   \title{Inclination effects in T Tauri star spectra}

   \author{Immo Appenzeller
          \inst{1}
          \and
          Claude Bertout\inst{2}
          }

   \institute{Landessternwarte, K\"onigstuhl 12,
              D-69117 Heidelberg, Germany\\
              \email{iappenze@lsw.uni-heidelberg.de}
         \and
             Institut d'Astrophysique, 98bis Bd. Arago, 75014 Paris, France\\
             \email{claude.bertout@obspm.fr}}

   \date{Received June 28, 2013; accepted July 25, 2013}

 
  \abstract
   {Because of the presence of rotation and accretion disks, classical T~Tauri
    stars have symmetry planes \textcolor{black}{that} are normally inclined relative to
    the plane of the sky. The inclination angles affect 
    the observed spectral properties of these objects.}
   {We study the influence of the inclination angles
   on classical T~Tauri star spectra in an empirical manner.}
   {Published inclination angles, 
   derived from the stellar photospheric rotation or from spatially
   resolved circumstellar disk observations, are compared with various
   observed spectral properties, and correlations are established and
   investigated.}
   {Inclinations derived from the stellar rotation are found to be much less accurate than the published disk inclinations, and no significant correlations between  spectral properties and inclinations based on rotation data could be detected. In contrast, significant correlations are found between the disk inclination
    angles and the apparent velocities observed for the forbidden 
    emission lines and the wind absorption features of permitted lines. These data support the assumption of cone-like polar winds with opening angles smaller than $\approx 45^\circ$.
    Other spectral features show weaker or no inclination dependence. 
    Using these results, the true (deprojected) flow velocities of the polar
    winds are derived for the investigated sample of T~Tauri stars. Deprojected wind-ejection velocities appear to differ by a factor of two among the stars in our sample, which  spans a range of mass-loss rates from $10^{-10} $M$_\odot$/yr to
$3 \times 10^{-7} $M$_\odot$/yr.}
   {}

   \keywords{stars: formation -- stars: pre-main sequence  -- stars: rotation -- stars: winds, outflows}
   \authorrunning{I. Appenzeller \& C. Bertout}
   \titlerunning{Inclination effects in T Tauri star spectra}
   \maketitle


\section{Introduction}
Classical T~Tauri stars (CTTSs) are low-mass pre-main sequence
objects with active accretion disks (e.g., Bertout, Basri \& Bouvier 
\cite{bertout88}). The accretion flow to the
stellar surface of CTTSs is known to take place along 
magnetic flux tubes that are anchored to the central stars
(e.g., Camenzind \cite{camenzind90}).
The theory of the formation of low-mass stars predicts that 
the rotational equatorial planes of the stars and 
the planes of the accretion disks are similar
(see, e.g., Yorke, Bodenheimer, and Laughlin \cite{yorke93}),
although the symmetry planes of the magnetic fields may differ
(e.g., Bouvier et al. \cite{bouvier07}; O'Sullivan et al. 
\cite{osullivan05}). Theoretical models 
of the star-disk systems and their line-emitting regions  
reproduce the basic observed properties of the CTTSs fairly well
(see, e.g., Hartmann, Hewett, and Calvet 
\cite{hartmann94}; Kurosawa, Romanova, and Harries \cite{kurosawa11}). 

However,
the details of the observed
spectra are expected to depend on the inclination angles of the individual
stars. In addition to the inclination-dependent rotational broadening of
the photospheric absorption lines, major effects on the emission 
line profiles and the strength and width of selected lines have been
theoretically predicted (e.g., Muzerolle, Calvet, and Hartmann 
\cite{muzerolle01}; Kurosawa, Harries, and Symington \cite{kurosawa06}).
But comparisons of the theoretically predicted inclination dependence with 
observed spectra have so far led to inconclusive results 
(e.g., Edwards et al. \cite{edwards94}; Kwan, Edwards, 
and Fischer \cite{kwan07}). 
Therefore, in the present study we identify inclination-dependent 
spectral features in CTTS spectra in a purely empirical way by comparing 
observed spectral properties with inclinations inferred from observations. 
As input data we use published empirically derived inclinations, 
which are based either on the rotational
properties of the corresponding stars or on spatially resolved observations 
of their circumstellar  disks.

\begin{table*}[]
 \caption[]{CTTSs used for this study, listed
following the order of the Herbig-Bell catalog of Orion population
stars (HBC).
}
\begin{tabular}{rrrrrrrrrrrrr}
 \hline \hline
  HBC &
  Object &
 irot &
 ref &
idisk & 
ref &
$V$be[OI] &
$V$b[NII] &
ref &
FWHM(H$\alpha $) &
FWQM(HeI) &
BAc(H$\alpha $) &
BAe(HeI)\\
 & & deg & & deg & & km s$^{-1}$
& km s$^{-1}$ & & km s$^{-1}$ \ & km s$^{-1}$ \ \ & km s$^{-1}$ \ & km s$^{-1}$ \
  \\  
\hline
25  & CW Tau   &    &   &    &    &-147 &-109 & 12 & 465 & 290 &-110 &  w  \\
26  & FP Tau   &    &   &    &    & -86 &     & 10 & 233 & 122 & -47 & -84 \\
28  & CY Tau   & 72 & 1 & 31 & 18 &-138 &     & 10 & 189 &  91 &   w &-161 \\
30  & DD Tau   &    &   &    &    & -79 & -80 & 10 & 195 &  69 & -50 & -58  \\
32  & BP Tau   & 32 & 2 & 36 & 18 & -31 &     & 10 & 315 &  70 &   w & w   \\
33  & DE Tau   & 57 & 2 &    &    &-148 &     & 10 & 300 & 125 &-55  & w   \\
34  & RY Tau   &    &   & 66 & 19 &-115 &     & 12 & 450 &   w &-40  &     \\
35  & T Tau N  & 15 & 1 & 20 &  6 &-180 &     & 15 & 275 & 225 &-190 &     \\
36  & DF Tau   & 50 & 2 &    &    &-103 & -84 & 12 & 295 & 75  &-63  &-163 \\
37  & DG Tau   & 58 & 2 & 32 & 18 &-276 &-219 & 10 & 225 & 285 &-200 &-383 \\
38  & DH Tau   & 58 & 1 &    &    & -50 &     & 12 & 167 &     & w   &    \\
43  & UX Tau A & 60 & 1 & 46 & 23 &     &     &    & 125 &   w &-155 & -95 \\  
45  & DK Tau   & 44 & 2 &    &    &-138 &     & 10 & 320 & 65  &-45  & -209 \\
48  & HK Tau B &    &   & 85 & 16 &-22  &-17  & 13 & 37  & 51  & w   &     \\
    & HH 30*   &    &   & 82 & 18 &-19  &-12  & 13 & 50  & 57  & w   &     \\
49  & HL Tau   &    &   & 40 & 24 &-198 &-171 & 13 & 239 & 275 &-148 &-300 \\
50  & XZ Tau   &    &   &    &    &-133 &-75  & 12 & 110 &     & w   &-177 \\
52  & UZ Tau E &    &   & 56 & 25 &-155 &-91  & 12 & 194 & 109 & -44 &-110 \\
54  & GG Tau   & 56 & 2 &    &    &-138 &     & 10 & 240 & 190 &-100 & w   \\
56  & GI Tau   & 54 & 1 &    &    &-69  &     & 10 & 165 &  68 & -45 & w   \\
57  & GK Tau   & 46 & 2 &    &    &-35  &     & 10 & 300 & 45  &-30  &-77  \\
58  & DL Tau   & 78 & 2 & 40 & 18 &-203 &-188 & 10 & 260 & 281 &-150 &-191 \\
60  & HN Tau   &    &   &    &    &-133 & -81 & 12 & 289 & 150 &  w  & w   \\
61  & CI Tau   &    &   & 46 & 25 &-179 &     & 10 & 228 & 202 & -113&-167 \\
63  & AA Tau   & 69 & 2 & 75 & 26 &-55  &-41  & 10 & 210 & 75  &-20  &-70  \\
65  & DN Tau   & 30 & 2 &    &    &-48  &     & 10 & 195 & 45  &-58  & w   \\
67  & DO Tau   &    &   &    &    &-121 &     & 10 & 200 & 58  &-100 &-158 \\
418 & HV Tau C &    &   & 84 & 17 &-41  &-25  & 13 & 98  & 109 & w   &     \\
72  & DQ Tau   &    &   &    &    &-52  &     & 10 & 240 & 115 & w   & w \\
74  & DR Tau   & 15 & 4 & 20 & 7  &-186 &-166 & 10 & 175 & 145 &-225 &-330 \\
75  & DS Tau   &    &   &    &    & -72 &     & 10 & 300 & 50  &-40  &-140 \\
76  & UY Aur   &    &   &    &    &-103 &     & 10 & 200 & 68  & w   &-135 \\
77  & GM Aur   & 90 & 1 & 52 & 18 &-19  &     & 10 & 315 & 140 &-75  &     \\
80  & RW Aur A & 37 & 2 &    &    &-221 &     & 10 & 490 & 125 &-75  &     \\
85  & GW Ori   & 30 & 5 & 10 & 7  &     &     &    & 180 &     & w   &-379 \\
250 & GQ Lup   & 27 & 3 &    &    &-38  &     & 22 & 297 & 67  &-13  &     \\
251 & RU Lup   &    &   & 28 & 7  &-171 &-173 & 11 & 258 & 220 &  w  &     \\
    & KH15D    &    &   & 84 & 14 &-35  &     & 14 & 121 &     & -7  &     \\
568 & TW Hya   & 18 & 9 &  8 & 8  &-12  &     & 20 & 190 & 129 &-45  &-260 \\
578 & VZ Cha   &    &   &    &    &-16  &     & 21 & 166 &  74 &  w  &     \\
590 & HM 32    &    &   &    &    &-14  &     & 11 & 173 & 58  &  w  &     \\
254 & AS 205 A &    &   & 47 & 25 &-237 &     & 11 & 183 & 269 &-180 &     \\
286 & S CrA N  &    &   &$<$45 & 7&-139 &     & 11 & 426 & 120 &-85  &     \\
291 & VV CrA   &    &   &    &    &-257 &-175 & 11 & 218 & 226 &-310 &     \\
292 & AS 353 A &    &   & 20 & 27 &-352 &     & 10 & 275 & 335 &-250 &-279 \\
\hline
\end{tabular}
\tablefoot{ The first column gives the HBC object number. 
The following columns give the most common object name, the 
inclination derived from the stellar 
rotation (irot) and the corresponding reference, the inclination derived 
from the disk aspect angle (idisk) and the corresponding reference, the 
radial velocity of the blue edge (measured at 25\% intensity) of the
630 nm [OI] line, the velocity of the blue peak of the [N~II] 658.3 nm 
line, the reference to the forbidden-line data, 
the H$\alpha $ FWHM line width, the FWQM width of the 587.6 nm HeI line, the 
central velocity of the blueshifted wind absorption feature in the 
H$\alpha $ emission
line (BAc(H$\alpha$)), and the blue edge (measured at 25\% depth) 
of the blueshifted wind absorption
feature of the 1083 nm He I line (BAe(HeI)). w indicates that the
corresponding feature could not be detected or measured.
}
\tablebib{(1)~Bouvier et al. \cite{bouvier95}; (2) Bouvier 2004, private  
communication; (3) Broeg et al. \cite{broeg07}; 
(4) Alencar et al. \cite{alencar01}; (5)
Shevchenko et al. \cite{shevchenko98}; (6) Ratzka et al. \cite{ratzka09}; 
(7) Schegerer et al. \cite{schegerer09};
(8) Akeson et al. \cite{akeson11}; (9) Alencar and Batalha \cite{alencar02}; 
(10) Hartigan et al. \cite{hartigan95}; (11) Hamann \cite{hamann94}; 
(12 ) Hirth et al. \cite{hirth97}; 
(13) Appenzeller et al. \cite{appenzeller05};
(14) Hamilton et al. \cite{hamilton03}; (15) B\"ohm and Solf \cite{boehm94}; 
(16) Stapelfeldt et al. \cite{stapelfeldt98}; (17) Stapelfeldt et al. 
\cite{stapelfeldt03}; (18) Guilloteau et al. \cite{guilloteau12}; 
(19) Isella et al. \cite{isella10}; (20) Herczeg et al. \cite{herczeg07}; 
(21) Krautter et al. \cite{krautter90}; (22) Appenzeller and Wagner 
\cite{appenzeller89}; (23) Tanii et al. \cite{tanii12}; 
(24) Kwon et al. \cite{kwon11}; (25) Andrews and Williams \cite{andrews07}; 
(26) Cox et al. \cite{cox13}; (27) Curiel et al. \cite{curiel97}.}
\end{table*}

\section{The database}
\label{database}   
This study is based on a sample of 45 CTTSs with well
studied spectra, for which direct or indirect
inclination information is available in the literature. A list of these stars 
and some relevant spectral data are given   
in Table~1. In 23 cases, accurate inclinations derived from spatially resolved
circumstellar disks are given. For 21 stars we include inclination 
values derived by comparing projected rotational velocity values $v \sin i$
with the rotational periods observed for the same stars. 
Since we found a tight correlation between disk inclinations and the 
velocities of strong forbidden lines, we supplemented our sample of stars with
published inclination data by 11 CTTSs with prominent forbidden 
line emission. In Table~1 these stars are listed without inclination
values. Of course, these stars could not contribute to establishing  
inclination-dependent correlations, but they allowed us to increase the
statistical significance of the correlations between the 
different inclination-sensitive spectral features discussed in 
Sect.~\ref{correlations}.    

Two of the 
stars without direct inclination information (GG~Tau and UY~Aur) are 
members of close multiple systems where 
the inclinations of circumbinary disks or rings are known. These 
inclinations are not listed in Table~1 because there are \textcolor{black}{close}
young binaries with
misaligned disks (Roccatagliata et al. \cite{rocca11}). Thus, inclinations
of the circumbinary disks may not be representative of the circumstellar
disk inclinations, which in both cases are known to be present 
in these systems. 
Also omitted in Table~1 are the published disk inclination values 
of the stars DN~Tau, DO~Tau, and DQ~Tau. Although inclination values 
for the circumstellar disks of these objects are 
quoted in the \emph{Catalog of Resolved Disks},
which is maintained by Karl Stapelfeldt\footnote{See http://circumstellardisks.org}, the disks of these three stars 
are only marginally resolved. Their inclinations are
therefore correspondingly uncertain. In one case (AS~353A), the disk 
inclination has
been derived indirectly from the well determined inclination of the
associated jet, assuming that the jet propagates in a 
direction \textcolor{black}{that is} normal to the disk
plane (as observed in all CTTSs for which this information exists).

For all inclination values listed in Table~1 the sources are given in the 
corresponding reference columns. Generally the source giving the
lowest observational error is quoted. Usually this is also the most recent
derivation. In many cases, earlier derivations 
are mentioned in the references listed in Table~1. Additional 
references on earlier inclination measurements can be found 
in the extensive literature lists of the \emph{Catalog of Resolved Disks}
cited above.

The references cited in Table~1 give mean errors of the listed disk 
inclinations between 1 degree and 6 degrees, with an average value 
of 2.2 degrees. Uncertainties of the 
disk inclination derivations of this order are
confirmed by comparing independent derivations by different authors 
and by measurements obtained at different wavelengths. 

The accuracy
of the inclinations derived from the stellar rotation is much lower.          
As described and discussed first by Weaver (\cite{weaver87}) the
rotational inclinations are derived by comparing projected 
rotational velocities $v \sin i$,
derived from photospheric absorption-line profiles, with the equatorial 
rotational velocities $v_{E}$. For $v_{E}$ we have
   \begin{equation}
   \label{equ1}
      v_{E} = 2 \pi R\, P^{-1},
   \end{equation}
where $R$ is the stellar radius, and $P$  the observed rotation period.
The inclination then follows from the relation 
    \begin{equation}
    \label{equ2}
      i = \arcsin[const. \times (v \sin i)\, P\, R^{-1}] = \arcsin Q,
   \end{equation}
where $Q = const. \times (v \sin i) P\, R^{-1}$.  
The error $\Delta i$ of the inclination angle
then becomes  
   \begin{equation}
      \Delta i = \frac{\partial i}{\partial Q} \Delta Q = (1-Q^{2})^{-1/2} 
   \Delta Q. 
   \end{equation}
For low values of $i$, the error of $i$ is approximately proportional to the
error of $Q$, but for $Q \rightarrow 1 $ (i.e. $i \rightarrow \pi/2$),
the error becomes very large.

Inclination estimates from $v \sin i$ are also prone to systematic errors
if additional line broadening effects are not properly taken into account. 
If additional broadening effects are overlooked, the evaluation 
of the line profiles tends to result in inclination angles that are 
systematically too large.

Of the observational input parameters of  Eq.~\ref{equ2}, the values
$v \sin i$ and $P$ can be measured fairly precisely. However,
the derivation of the stellar radius $R$ depends on parameters, such as 
the luminosity, the 
effective temperature, the reddening, and the veiling of the stars, which
are often not very well known for CTTSs. Therefore, stellar radii of CTTSs 
derived by different
authors typically differ by factors of two (see, e.g., Johns-Krull and
Gafford \cite{jkrull02}). Even with $R$ being uncertain by that
amount, for small $i$ a useful approximation of the inclination can still be 
estimated from stellar rotation data. But for inclinations $i > 30$ 
degrees, the errors of $i$ derived from the rotation 
can \textcolor{black}{approach} $\pi/4$. In this case the inclination essentially becomes undetermined.       

In addition to observed inclinations, we present in Table~1 velocity
data derived from the profiles of selected emission lines. 
For the forbidden lines, we provide individual references in the column
following the [N~II] data. The mean errors of the forbidden line 
velocity data range between 2~km/s and about 10~km/s, with an 
average of about 6~km/s. The information on the profiles of
the H$\alpha $ and He~I~587.6~nm lines was in most cases taken from 
Alencar and Basri (\cite{alencar00}). For the objects not observed by these
authors, the following sources were used: Beristain, Edwards, and Kwan 
(\cite{beristain01}) for FP~Tau, CY~Tau, DD~Tau, GI~Tau, and HN~Tau;
Appenzeller, Bertout, and Stahl (\cite{appenzeller05}) for HK~Tau~B, HH30*,
and HV~Tau~C; Krautter, Appenzeller, and Jankovics (\cite{krautter90}) for VZ~Cha and S~CrA; 
Appenzeller and Wagner (\cite{appenzeller89}) for GQ~Lup; Alencar and Batalha
(\cite{alencar02}) for TW~Hya; Hamann and Persson (\cite{hamann92})
for HM~32; Appenzeller, Jankovics, and Jetter (\cite{appenzeller86}) for VV~CrA;
Hirth, Mundt, and Solf (\cite{hirth97}) for H$\alpha $ of~DD Tau, DH~Tau, and HN~Tau;
Boesgaard (\cite{boesgaard84}) for H$\alpha $ of RU~Lup; Hamilton et al. 
(\cite{hamilton03}) for H$\alpha $ of KH15D; Stempels and Piskunov 
(\cite{stempels02}) for He~I~587.6~nm of RU~Lup.
The edge velocities of the He~I~1083~nm absorption feature in the last column
of Table~1 are based on the
profiles published by Edwards et al. (\cite{edwards06}). When 
spectra were not available in digital form, the corresponding 
features were measured in the published printed figures. 
The resulting accuracy (m.e. $\approx $ 10 km/s) is adequate     
for the statistical purposes of this study. For the blueshifted [N~II] 
emission peak and the wind absorption feature of the H$\alpha $
line, we present central velocities. For [O~I]~630~nm (where the
the blueshifted peak is often not well defined) and for the He~I~1083~nm absorption feature, we give edge velocities, as defined in the
notes to Table~1. This has to be taken into account when
the velocities are compared.

\begin{figure}
\label{F1}
\centering
\includegraphics[angle=0,width=10cm]{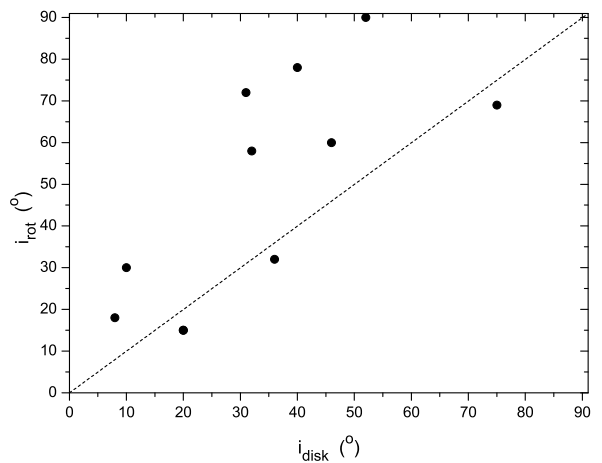}
\caption{Inclination angles $i_{rot}$ derived from the stellar rotation 
as a function of the circumstellar disk inclinations $i_{disk}$ of 
the corresponding objects.}
\end{figure}

\section{Comparison of rotational and disk inclinations}
\label{rotdisk}
The spectral data given in Table~1 were used to search for 
possible correlations between the inclination and spectral
properties. Initially this search was carried out separately for
the inclinations based on the stellar rotation and for the inclinations of
the circumstellar disks. However, for the inclination angles derived from 
the rotational data, no significant correlations could be 
detected for any of the tested spectral features. On the other hand,
correlations were found, when the same
spectral features were compared with the disk inclinations. 

Therefore,
in a next step,  we used the 11 objects in Table~1 for which
inclinations derived from $v \sin i$ and disk inclinations
are both available to compare the inclinations derived by
the two methods.
As illustrated by Fig. 1,
the inclinations derived by the two methods are 
correlated for this sample, but are clearly not identical. The mean difference between the
inclinations derived by the two methods is about $19^\circ$. More critical
is that the inclinations derived from $v \sin i$ are 
systematically larger (by about the same amount). In three cases with
$i > 30^\circ,$ the difference is about $40^\circ$, i.e., close to
$\pi/4$. The average differences
between the inclination angles derived by the two methods greatly exceed    
the observational errors given for the disk inclinations, but 
(in view of the errors discussed in Sect.~\ref{database}) such deviations 
are expected 
for the rotation-based inclinations. Together with the fact that no 
correlations between the rotation-based inclinations and the spectral 
properties could be detected, we regard the poor agreement between 
the angles indicated in Fig.~1 as evidence that the 
rotation-based inclinations are, at present, not 
accurate and reliable enough for
statistical studies. Therefore, the following sections will be  based
essentially on the disk-inclination values.
     
\begin{figure}
\label{F2}
\centering
\includegraphics[angle=0,width=10cm]{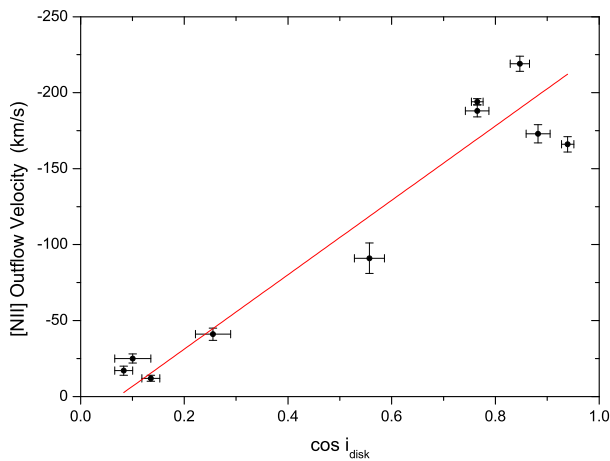}
\caption [] {Projected wind velocities inferred from the blueshifted peaks 
of the [N~II] 658.3 nm lines as a function of the cosine of
the disk inclination. Also included is the corresponding linear 
regression line.}
\end{figure}

\section{Correlations between the disk inclinations and spectral properties}
\label{correlations}
Among the spectral features for which a correlation with inclination
has been predicted most frequently are the emission lines and 
absorption features that 
are formed in the winds of the CTTSs. This applies 
in particular to the broad or high-velocity component of the 
forbidden-line profiles (see, e.g., Hartigan, Edwards, and Ghandour 
\cite{hartigan95}) and to the blueshifted absorption components 
of the hydrogen and helium permitted lines (e.g., Kurosawa et al. 
\cite{kurosawa06}; Beristain, Edwards, and Kwan \cite{beristain01}). 
In the following, we investigate in turn the forbidden and 
permitted line profiles.

\subsection{Forbidden lines}
The forbidden lines in CTTS spectra often have complex 
profiles, which
vary greatly between individual objects and between different lines.
However, in many cases the profiles can be deconvolved into a narrow
or low-velocity component (LVC) with a typical FWHM of about 50~km/s
(see Hirth et al. \cite{hirth97}), and a broader and usually blueshifted high-velocity component (HVC). 

The \textcolor{black}{HVCs} are generally assumed to 
originate in fast outflows, which continue into the 
interstellar space as jets. Among the lines where the broad 
forbidden-line component is usually most conspicuous are the [N~II] 
lines at 654.8~nm and 658.3~nm. In many cases, the [N~II] 
line profiles have a well
defined blueshifted peak. The absolute values of the peak velocities 
are usually regarded as a mean for the projected wind velocities 
of the [N~II] emitting volumes. If available, the measured 
velocities of the blue [N~II]~658.3~nm peaks are listed in Table~1. 

At high inclinations, the blueshifted peak
tends to merge with the low-velocity peak, and for edge-on objects the
blueshifted peak can often not be separated from the narrow, 
low-velocity component. Among the objects in Table 1, this applies to
HK Tau B and HH 30*. In these cases we include in Table~1 the highest 
value of the HVC-peak blueshift \textcolor{black}{that} appears to be compatible with the blue wing of the total [N~II] profile. Since the corresponding velocity limits are close to zero (but - by definition - cannot be $> 0$), they are,
within the accuracy of this study, adequate approximations of the
blueshifted  peak velocities. 

Resolved jet flows of CTTSs often form narrow and almost linear cones. 
Assuming that the unresolved inner [N~II] flows have the same geometry, we 
may expect that the observed wind velocities are proportional to $\cos i$.
Therefore, in Fig. 2 we plot the projected wind 
velocities, as inferred from the velocities of the blue [N~II] 
peaks, as a function of $\cos(i_{disk})$.
The figure shows that the observed [N~II] wind velocities are clearly
correlated with the disk inclinations. The correlation's adjusted coefficient of determination (which takes the error bars into account) is $\bar{R}$~=~0.95. Moreover, the plot supports 
the assumption that the $\cos i$ -law provides a
good approximation. Most of the variation in the observed
velocity can obviously be explained by the varying inclination.            

\begin{figure}
\label{F3}
\centering
\includegraphics[angle=0,width=10cm]{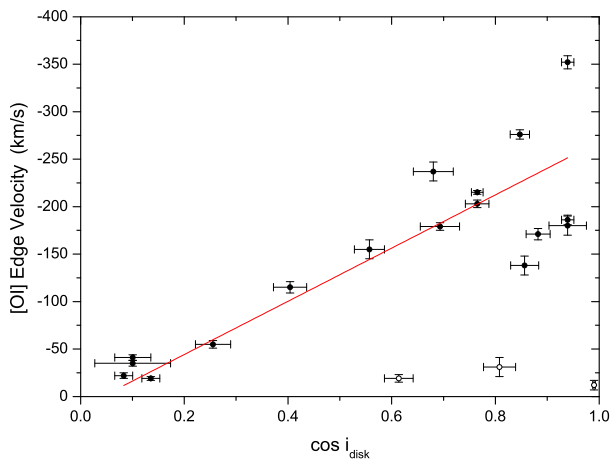}
\caption{Projected wind velocities inferred from the blue edges
of the [O~I]~630~nm lines as a function of the cosine of the disk inclination 
and the corresponding linear regression line. The open symbols refer to
the three ``weak-wind'' objects BP Tau, GM Aur, and TW Hya.}
\end{figure}

The projection effect is also expected to influence the apparent velocity
observed in the [O~I]~630~nm line, which forms under similar conditions 
to the [N~II] line. The [O~I] line tends to be stronger and is 
present in the spectra of practically all stars listed in Table~1. However,
compared to [N~II], the oxygen line has the disadvantage of often lacking a
distinct blue peak, although a broad line component is usually 
present in the form of an extended blue wing. As a measure of the projected (maximal) wind velocity, for this line we 
therefore used  the blue edge 
of the total line profile, measured at 25\% intensity. 
In Fig.~3 this quantity is plotted as a function of $\cos(i_{disk})$. 
Again a clear correlation is observed. However, three objects, 
BP~Tau, GM~Aur, and TW~Hya
fall well below the relation defined by all other stars. Therefore, these
three stars are discussed separately in Section \ref{WWS}.
If we disregard the three objects listed above,
the correlation in Fig.~3
is again highly significant, with $\bar{R}$~=~0.91.            

Since both the blue-peak velocity of the [N~II] profile and the [O~I] 
blue-edge velocity have been found to depend on the inclination, there  
has to be some mutual correlation between these two quantities.
This is illustrated in Fig.~4. Since the [O~I] edge measures the maximal
outflow velocity, it is expected to be systematically higher than the [N~II] peak velocity. This is confirmed by Fig.~4, where the 
[O~I] edge velocities are on average 25\% higher 
than the [N~II] peak velocities. The individual 
points in Fig.~4 deviate from the mean relation on average 
by only about 20~km/s. This value
is not much more than the differences expected from the measuring
errors alone ($\bar{R}$~=~0.96 for this correlation).

\subsection{Weak-wind stars}

\label{WWS}
As noted above and shown in Fig.~3, the three stars BP Tau, GM Aur, and TW Hya
were found to have [O~I] absolute edge velocities~$\leq$~31~km/s
in spite of $\cos i$ values~$>$~0.6. In all three cases, the observed [O~I]
edge velocity differs from the mean relation by more than 20 $\sigma $.
All three stars have only very weak forbidden lines without 
detectable broad components. In contrast to the edge-on  CTTSs with narrow
[O~I] lines, these three stars have no detectable jets (e.g., 
Azevedo et al. \cite{azevedo07}). Moreover, except for the blueshifted 
absorption component of the He~I~1083~nm line of TW~Hya,  
wind absorption features of the permitted lines
are very weak or absent in the spectra of these stars. 
GM~Aur and TW~Hya are relatively evolved
pre-main-sequence stars with estimated evolutionary ages of, respectively, 
7.4~Myr and $\approx $10~Myr (Bertout et al. \cite{bertout07}, Webb et al.
\cite{webb99}).  In the spectrum of GM~Aur, 
the emission lines sometimes almost disappear. During such 
periods, GM~Aur has the spectral properties of a  
weak-line TTS (Edwards et al. \cite{edwards06}). 
BP~Tau has a lower age (3.2 Myr, according to Bertout et al. \cite{bertout07}).
But, according to Dutrey et al. \cite{dutrey03}, this star is ``a transient
object in the process of clearing its disk''. Kwan, Edwards, and Fisher
(\cite{kwan07}) tentatively explained the absence of wind absorption 
features in the spectrum of BP~Tau by an inclination effect. However,
as shown later in this section, other stars with similar or higher 
inclinations do have wind absorption components, and an inclination effect
cannot explain the weak and narrow forbidden lines of BP~Tau. 

In view of the low observed  [O~I] edge velocity and the weak wind absorption features BP~Tau, GM~Aur, and TW~Hya, which in the following are 
denoted as ``weak-wind stars'', are indicated \textcolor{black}{by open symbols} in the
following plots, and they are not included in graphs showing
deprojected wind velocities.

\begin{figure}
\label{F4}
\centering
\includegraphics[angle=0,width=10cm]{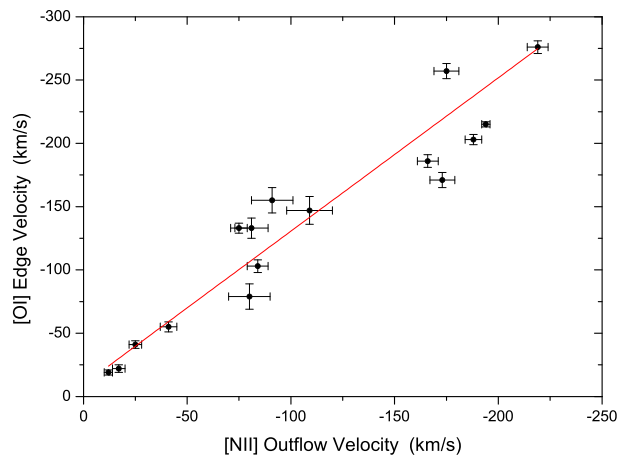}
\caption{Blue-edge velocities of the 
of the [O~I]~630~nm lines as a function of the projected wind
velocities indicated by the blue peak of the [N~II]~658.3~nm 
line, and the corresponding linear regression line.}
\end{figure}

Some other stars in Table~1 may also belong to the weak-wind category,
although a confident assignment to this class is only possible for the 
three objects mentioned above, for which reliable inclination data are 
available. The most likely additional candidate 
for the weak-wind group in our sample is the CTTS GQ~Lup, 
which is one of the best studied TTSs. High-resolution spectra of this
object cover 
more than 35 years. All show a typical CTTS spectrum 
with significant mass accretion. However, the star again belongs to 
the more evolved TTSs (e.g., Donati et al. \cite{donati12}, Johns-Krull 
et al. \cite{jkrull13}). No reliable disk inclination has been
reported for GQ~Lup. But its rotation-based inclination ($27\,^{\circ}$) is 
low and therefore probably not dramatically wrong (see Sect. \ref{rotdisk}). 
In spite of its likely low inclination, GQ~Lup has
a very low blueshift of its [O~I] emission edge, and its [O~I] 630 nm
emission strength (EW~$\approx 80$ m\AA ) is the lowest of all
objects in Table~1.

\subsection{Wind absorption features of permitted emission lines}

\begin{figure}
\label{F5}
\centering
\includegraphics[angle=0,width=10cm]{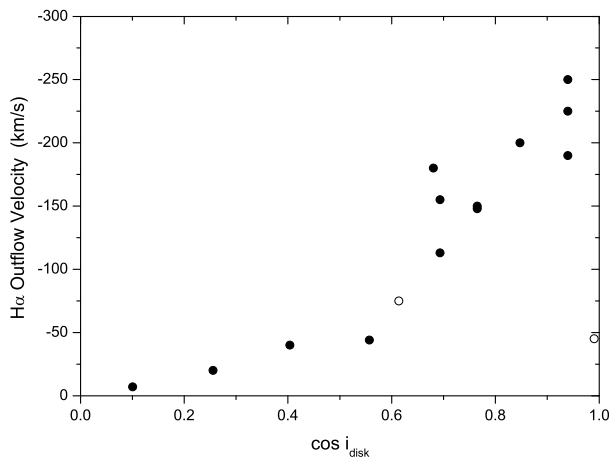}
\caption{Projected wind velocities inferred from the blueshifted absorption
component of the H$\alpha $ line as a function of the cosine of the
disk inclination. Because of the time variability of the H$\alpha $ profiles,
errors cannot be reliably estimated for the values given for this feature.
Therefore, error bars are omitted in this plot. The open symbol refers to
GM Aur. Note the weak dependence on $\cos i$ of the outflow velocity 
for $\cos i < 0.7$.  
}
\end{figure}

\textcolor{green}{\textcolor[rgb]{0,0,0}{Other spectral features usually ascribed t}}o the winds of CTTSs are the 
blueshifted absorption components of strong permitted emission lines.
Thus, their observed velocities are expected to be correlated
with the inclination, too. As shown in Fig.~5, the central velocity of the
blueshifted absorption component in the blue wings of the
H$\alpha$ lines does, as expected, increase with $\cos(i_{disk})$.
However, the relation observed in Fig.~5 clearly
differs from \textcolor{black}{the one} found for the forbidden lines, as plotted in
Figs.~2 and~3. While for the forbidden lines
the apparent velocity increases more or less linearly with $\cos i$,
in Fig.~5 the projected velocity remains low for approximately 
$\cos i < 0.7$ (corresponding to $i > 45^\circ$) and increases
steeply for smaller inclinations. Since an absorption feature requires 
that the line-of-sight passes through the absorbing volume, as well as through
the emission source, such behavior is to be expected for a wind
that is restricted to a polar flow cone (see, e.g., Edwards et al. 
\cite{edwards06}). For a point-like continuum emission source, Fig.~5 
can be explained by a polar flow cone with an opening angle of $45^\circ$. 
However, depending on the relative sizes and distance of the absorbing and continuum emitting volumes, Fig.~5 would also be consistent with a much
smaller opening angle and an extended continuum emission region.

Among the permitted lines in TTS spectra, the He~I~1038~nm line, which has a metastable lower level, is particularly well suited   
for studying the kinematics of absorbing gas flows.       
Therefore, we compare in Fig.~6 the apparent wind velocities
inferred from this line with the observed disk inclinations. Since
the absorption by this line usually covers a broad velocity 
range, and since we are mainly interested in the terminal velocity of the
wind, we use the blue-edge velocity of the absorption 
component for this line. To have a consistent and reproducible measure, we  
plot the velocity where the blue wing of the wind absorption 
reaches 25\% of the central depth in the profiles published by 
Edwards et al. (\cite{edwards06}). The sample for which we
have data on this spectral feature, as well as on disk inclinations,
only contains one object with $\cos i < 0.5$. Therefore, we use the 
disk inclination angle in Fig. 6, rather than its cosine, as the abscissa. 
The figure again demonstrates that there is a correlation. 
The data points are consistent with a steep increase in the apparent
wind velocity  for approximately $i < 45^\circ$ (as observed for the H$\alpha $ wind 
absorption), but the lack of more data at high inclinations does not allow us
to prove this behavior independently \textcolor{black}{in the case of} the He I wind absorption \textcolor{black}{alone}.
Interestingly, the He~I edge velocity of the weak-wind star TW~Hya 
(-260 km/s, $i_{disk}$ = $8^\circ$)
also fits the trend shown in the figure. (For the other weak-wind objects,
the corresponding spectral feature is very weak or absent.) 

\begin{figure}
\label{F6}
\centering
\includegraphics[angle=0,width=10cm]{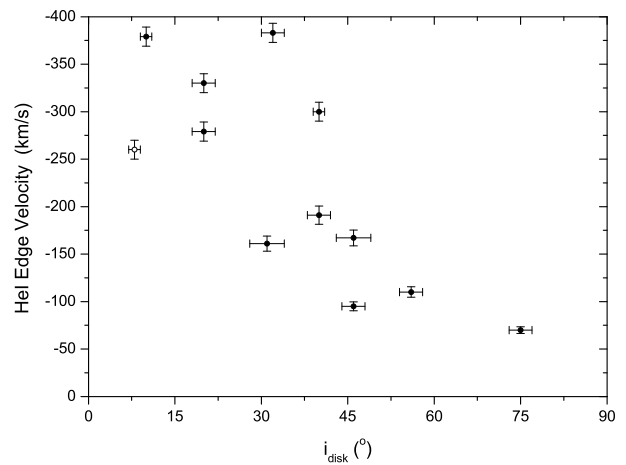}
\caption{Projected wind velocities inferred from the blue edge 
of the wind absorption component of the He I 1083 nm 
line as a function of the disk inclination. The open symbol refers to
TW Hya.}
\end{figure}

\begin{figure}
\label{F7}
\centering
\includegraphics[angle=0,width=10cm]{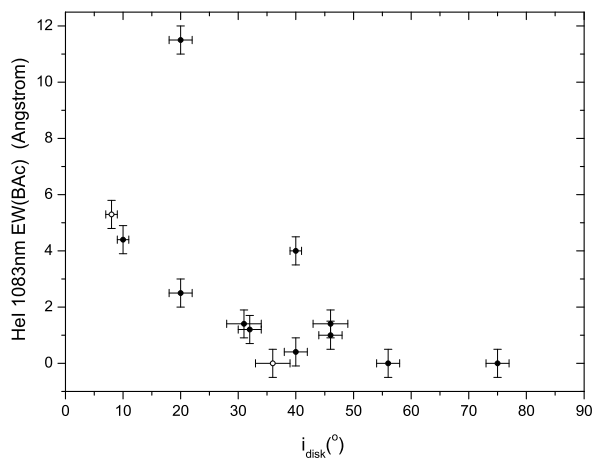}
\caption{Equivalent width below the continuum of the blueshifted 
absorption component of the He~I~1083~nm line profile as a 
function of the disk inclination. The two open symbols refer to
BP Tau and TW Hya.}
\end{figure}

A related correlation with the disk inclination is shown in Fig. 7,
where we plot the equivalent width below the continuum of the 
blueshifted absorption component of the
He~I~1083~nm line. The corresponding data are again taken from 
Edwards et al. (\cite{edwards06}). The uncertainty of the EW values has 
been estimated from the scatter of those cases where observations 
for more than one epoch have been reported. The absence of a 
significant absorption for
inclinations $> 45^\circ$ again supports the assumption of a cone-like 
outflow geometry with an opening angle $\leq 45^\circ$.
The data point of the weak-wind star TW Hya 
(EW~=~5.3~\AA , $i_{disk} = 8^\circ$) again fits the trend indicated 
in Fig.~7. 

Since the velocities of the wind absorption features and those of the 
forbidden lines were both found to be correlated with the inclination,
we also expect correlations between the forbidden-line velocities
and the wind absorption.  This is confirmed in Figs.~8 and 9, 
where we plot our measures for the wind velocities of the permitted 
lines as a function of 
the [O~I] outflow velocities. Interestingly, for the He~I absorption
edge, we obtain \textcolor{black}{on average} fairly similar projected outflow velocities to those for
the [O~I] emission line edge, although these features can hardly be produced
in the same volumes. The central velocities of the H$\alpha $ absorption features have \textcolor{black}{on average} only about two thirds of the [O~I] edge velocities.
Although at least part of the correlations indicated in Figs.~8 and 9 certainly result from an inclination effect, a contribution 
\textcolor{black}{from} a correlated intrinsic wind velocity cannot be excluded. But in
both cases it is clear from the figures that the winds seen in \textcolor{black}{both} the
forbidden lines and  the permitted lines are physically related,
in spite of the very different conditions for \textcolor{black}{ forming} these
different spectral features.  

\subsection{The He~I~587.6~nm line width}
\begin{figure}
\label{F8}
\centering
\includegraphics[angle=0,width=10cm]{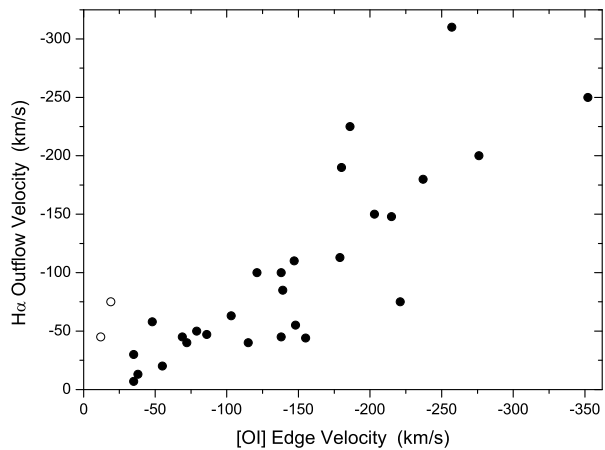}
\caption{Projected wind velocities inferred from the blueshifted  
absorption component of the H$\alpha $ line as a function of the 
projected wind velocity indicated by the blue edge of the
[O~I] 630 nm line. The open symbols refer to GM Aur and TW Hya.}
\end{figure}

\begin{figure}
\label{F9}
\centering
\includegraphics[angle=0,width=10cm]{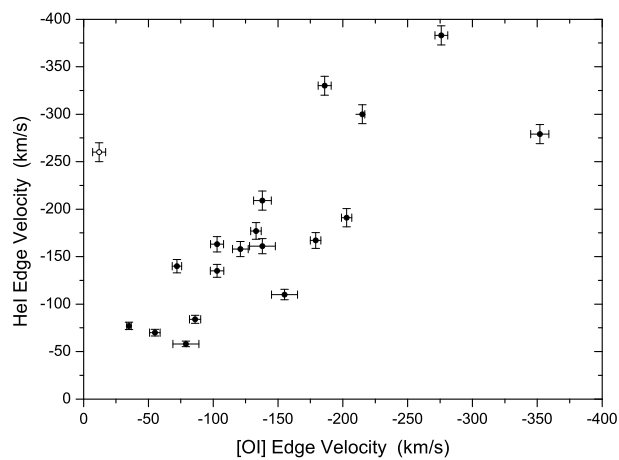}
\caption{Projected wind velocities inferred from the blue edge of 
the wind absorption component of the He~I~1083~nm line as a function of  
the projected wind velocity indicated by the blue edge of the
[O~I]~630~nm line. The open symbol refers to TW Hya.}
\end{figure}

\begin{figure}
\label{F10}
\centering
\includegraphics[angle=0,width=10cm]{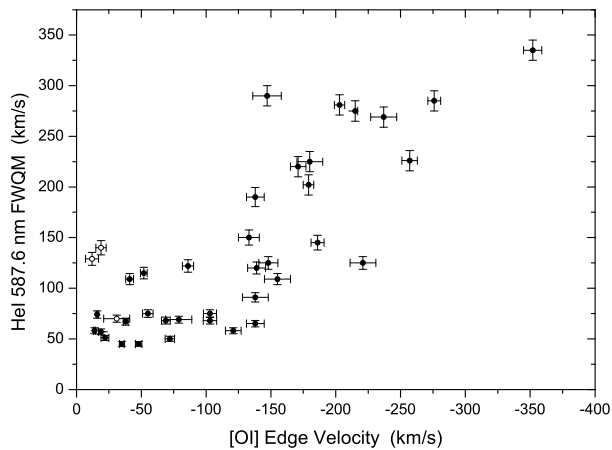}
\caption{Full width at one quarter of the maximum of the 
He~I 587.6 nm emission line as a function of the  
projected wind velocity indicated by the blue edge of the
[O~I] 630 nm line. The three weak-wind objects are again indicated by
open symbols.}
\end{figure}

Another permitted line that has been discussed extensively in the context 
of the polar winds of CTTSs is the He~I~587.6 nm emission feature
(see, e.g., Beristain, Edwards and Kwan \cite{beristain01}; Kwan and
Fischer \cite{kwan11}). Like the forbidden lines, this He~I line often 
\textcolor{black}{consists of both} a broad and a narrow component. The broad components 
of the He~I~587.6~nm line have widths that are comparable   
to those of the broad components of the forbidden 
lines. However, the profiles of the broad components of the
helium lines are quite different from those of the broad components of the
forbidden lines. While the forbidden lines often have a steep blue wing
and no red wing, the broad components of the helium lines are
in most cases essentially symmetric with a Gaussian-like profile.  
The narrow components of the helium lines are also symmetric
and slightly broader than the
corresponding forbidden line features. Also in contrast to the corresponding
forbidden-line broad components, the broad \textcolor{black}{helium-line} components are   
only modestly blueshifted relative to the systemic velocity by 
about 75~km/s on average. 

The broad \textcolor{black}{helium emission} component is often assumed to form in the 
T~Tauri winds (e.g., Kwan and and Fisher \cite{kwan11}). 
Therefore, we compared the
observed widths of the He~I~587.6~nm line with the disk inclinations. As
a measure of the width, we used the full width of the line at a quarter 
of the maximal intensity (FWQM) of the total line. For the 15
objects where disk inclination data and the He~I~587.6~nm line
widths are listed in Table~1, we find a decrease in the
line width with increasing inclination, which is to be expected when the width
is related to a polar wind. However, the scatter around the mean
relation is too large to regard this result as conclusive. 

On the other hand, a more convincing  
correlation is found when the FWQM of the total  He~I emission
is compared with the [O~I]  edge velocity (Fig.~10). Although the scatter
of the individual data points is large, the relatively high number of  
data points makes the correlation significant.
Since (as shown above)
the [O~I] velocity is well correlated with the inclination, Fig.~10 
may provide indirect evidence for a relation between
the maximal velocity observed in the He~I 587.6 nm line\footnote{or some other measure of the
broad-component width of the He~I~587.6~nm line, such as the parameter
$V_{bwing}$ listed by Beristain, Edwards, and Kwan (\cite{beristain01}).} and the
inclination, which    
supports a wind origin of at least part of this line emission. But the sample for which
these informations \textcolor{black}{are} available is
too limited to derive firm results. No significant correlations between 
the inclination and the FWHM of the broad component of the He~I line
(again listed in Beristain, Edwards, and Kwan \cite{beristain01}) 
and no correlation between the inclination and 
the intensity ratio between broad and narrow He~I line 
components (BC/NC) could be detected, in contrast to the result 
for the forbidden lines. However, we suspect that
such correlations do exist and could be established with 
larger samples.  

A closer look at the individual lines reveals a qualitatively 
different  dependence on inclination for the forbidden-line 
and helium-line profiles. For the forbidden 
lines, the broad component simply becomes narrower with increasing
inclination and finally merges into the narrow component. In the case of
the helium lines, broad components are found at all inclinations,
although their widths and strengths decrease  
with increasing inclination. A good example for the different 
behaviors of the forbidden lines and the
helium lines is the spectrum of the edge-on object HH30* (i~=~$82^\circ$),
which shows a weak broad component of the He~I~587.6~nm line 
(FWHM $\approx 140 $~km/s ), while a broad component of the [O~I]~630~nm
line is not detectable (Appenzeller, Bertout, and Stahl 
\cite{appenzeller05}). Similar behavior is seen in the spectrum of 
AA Tau (i~=~$75^\circ$).

\section{Negative results}
In addition to those discussed in Sect.~\ref{correlations}, we 
searched for correlations between the disk inclinations and various 
other spectral properties, such as the strength and width of
selected emission lines or line components, the veiling, and
the presence and strength of redward-displaced (accretion) absorption 
features. Apart from some quantities that are related to the
line profile data discussed in Sect.~\ref{correlations}, no other
significant correlations were detected. In contrast to a tentative 
result noted earlier (Appenzeller, Bertout, and Stahl \cite{appenzeller05})
and expectations from the theory (Kurosawa, Harries, and Symington
\cite{kurosawa06}), no significant correlation could be found
between the H$\alpha $
equivalent width and the inclination in our larger sample with  
more reliable inclination angles. If such a 
relation exists, it is obliterated by the very large 
scatter in the equivalent widths of the individual stars and by the
variability of this line. Larger samples will probably be 
needed to detect a corresponding correlation by statistical means. 

A search for a correlation between the H$\alpha $ FWHM line width and the inclination also ended inconclusively. Although the line widths were found to be on average  lower for disk inclinations $> 50^\circ$, the scatter of the individual values is very large, and the result is far from being 
statistically significant.

We did not detect any correlation between the disk inclination and
the presence or strength of the redshifted (``YY~Orionis'') absorption
components of the He~I~1083~nm line profiles observed by Edwards et al.
(\cite{edwards06}) and Fisher et al. (\cite{fischer08}). 
Redshifted absorption components were found at all 
disk inclinations between $8^\circ$ and $75^\circ$, and no dependence  
of their strength \textcolor{black}{on} the inclination is evident from the available 
data. The lack of such a correlation may be explained by the dependence
of the accretion flows on the
geometry of the magnetic fields, which in many cases is known to
be tilted relative to the rotation axis (e.g., Bouvier et al. 
\cite{bouvier07}; O'Sullivan et al. \cite{osullivan05}). Moreover, 
the redshifted absorption
components are known to show strong time variations, which may mask
any correlation with $i$. 
 
\begin{figure}
\label{F11}
\centering
\includegraphics[angle=0,width=10cm]{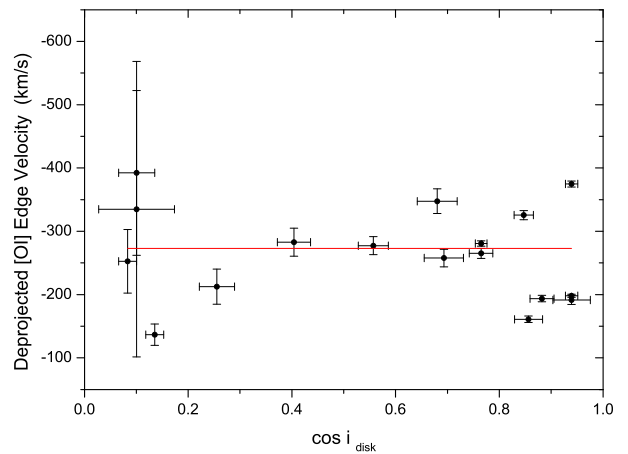}
\caption{Deprojected outflow velocities, as
derived from the blue edge of the [O~I] emission line,
as a function of the cosine of the disk inclination.}
\end{figure}

\section{Deprojected wind velocities} 
Assuming that the polar outflows from the CTTSs are normal to the disk
planes, one can derive the true, deprojected flow velocities by dividing  
the observed velocities by the cosine of the corresponding disk inclination, as
listed in Table~1. We 
carried out these calculation for the [O~I], [N~II], H$\alpha$, and He~I
velocities where disk inclination data are available in Table~1. 
Figure~11 gives the resulting deprojected [O~I] edge velocities 
as a function of $\cos i$. For the reasons
outlined in Sect.~\ref{correlations}, the three weak-wind 
TTSs have been omitted in this figure. The large error bars near the left
margin of the figure result from the $(\cos i)^{-1}$ magnification
of all errors for high inclinations. As shown by Fig.~11, the deprojected 
velocities scatter around a horizontal line. No systematic 
dependence of the deprojected velocities on the inclination
is detected, which appears to confirm the correctness of the
deprojection procedure. An error-weighted least square solution results 
for the complete sample of Fig.~11 in a deprojected flow velocity of
$273 \pm 14$ km/s. A simple average of all deprojected velocities of  
Fig.~11 gives a flow velocity of $262 \pm 18$ km/s. An average of 
the (possibly 
more reliable) data for the 13 objects with $i < 80^\circ$ ($\cos i >0.17$) 
gives $258 \pm 18$ km/s. 
The median deprojected velocity of both samples is 258~km/s. All deprojected
velocities range between 137~km/s $< v_{depr}< $ 392~km/s. If we 
disregard the less certain values for $i >80$ degrees, the range narrows to
the interval 161~km/s to 374~km/s. Neglecting the results for 
HV~Tau~C and KH15D, which have
errors of the deprojected velocities exceeding 100~km/s, we find that
80 \% of all flow velocities are in the range between 190~km/s and 350~km/s. 

\begin{figure}
\label{F12}
\centering
\includegraphics[angle=0,width=10cm]{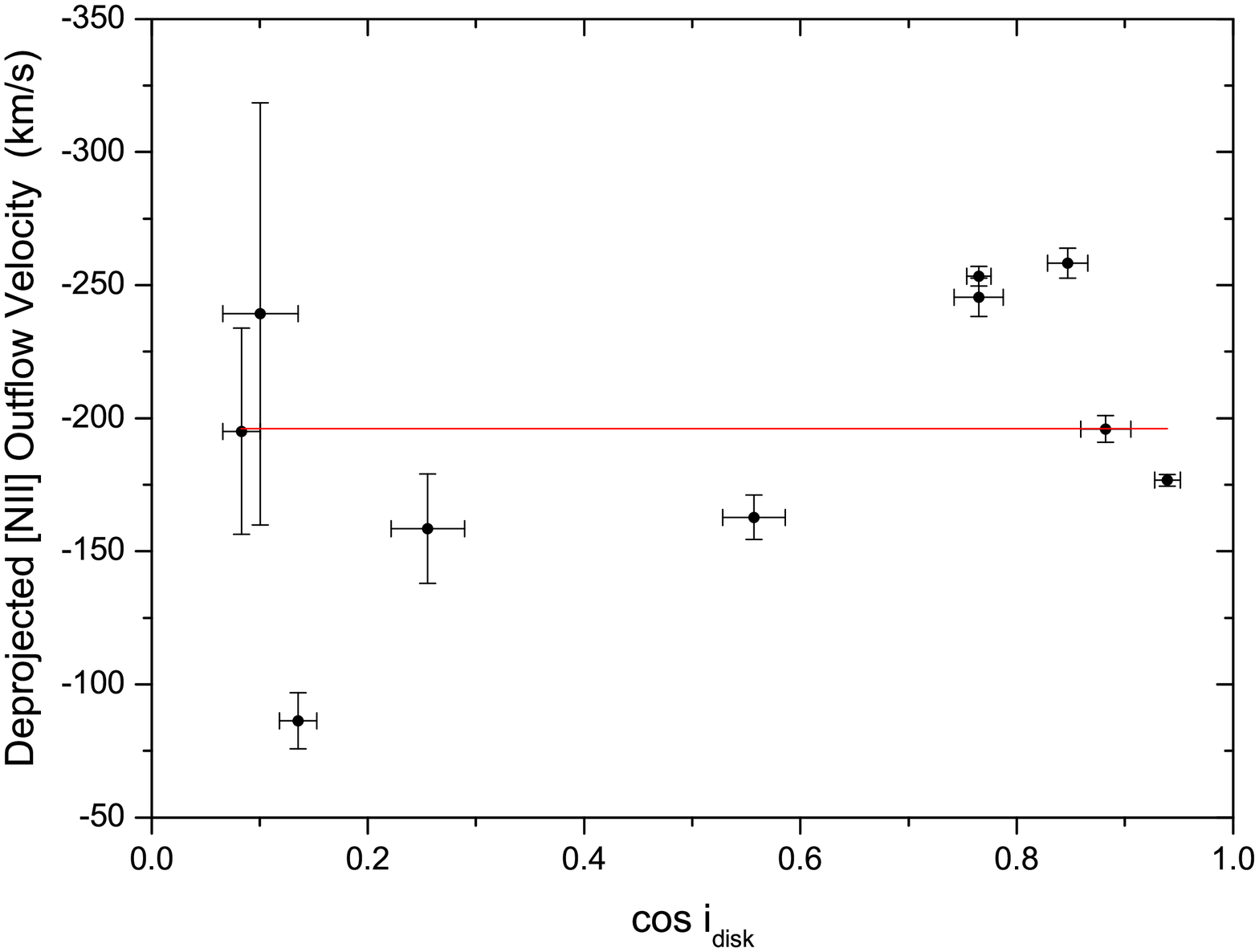}
\caption{Deprojected outflow velocities, as
derived from the blue peak of the [N~II] emission lines,
as a function of the cosine of the disk inclination.}
\end{figure}

Deprojected velocities of the [N~II]
blue emission peaks are plotted in Fig.~12. Because these values
correspond to to a mean flow velocity (while the [O~I] data correspond 
approximately to 
the maximal flow velocity), the [N~II] velocities are on average lower by about
20\%. For the [N~II] line, the error-weighted mean and the simple average
for all objects is $196 \pm 16$ km/s in both cases. If only 
the objects with $i < 80$ degrees are averaged, we get $205 \pm 14$ km/s.
The median value is 195.5 km/s. 

\begin{figure}
\label{F13}
\centering
\includegraphics[angle=0,width=10cm]{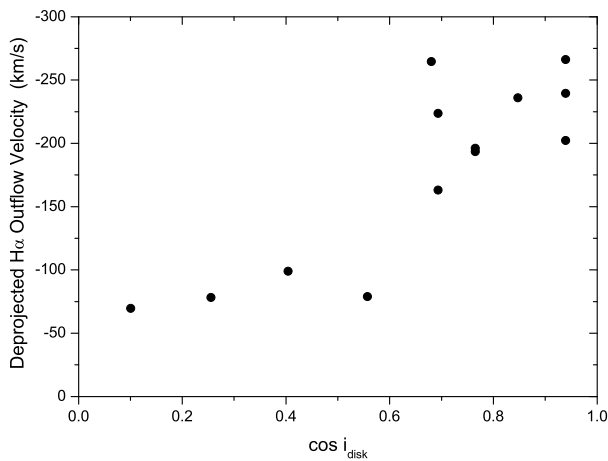}
\caption{Deprojected outflow velocities indicated by the blueshifted
absorption components of the H$\alpha$ line profiles as a function of the
cosine of the disk inclination.}
\end{figure}

In Fig.~13 we give deprojected velocities of the blue absorption
component of the  H$\alpha$ lines. As discussed in \textcolor{black}{Sect.}~\ref{correlations},
the velocities of this spectral feature behave differently for inclinations
below and above about $45^\circ$. This bimodality is confirmed by 
Fig.~13, which indicates a low deprojected velocity for
high inclinations ($\cos i \ < 0.68$) and high velocities for smaller
inclinations. The pattern is again consistent with the assumption of
a fast polar wind with a cone-like geometry and an opening 
angle $\leq 45^\circ$. The mean flow velocity derived from the H$\alpha$ blue 
absorption components with $\cos i_{disk} > 0.68$ is $224 \pm 11$ km/s,
which is slightly higher than the average [N~II] (peak) flow 
velocity, but lower than the mean deprojected [O~I] flow velocity.

Deprojected flow velocities were also calculated for the blue-edge 
velocities of the
He~I~1083~nm absorption components. 
The scatter of the resulting data with $\cos i > 0.68$ is
larger than in the case of the [O~I] emission line edges, but the
mean value $292 \pm 33$ km/s is (within the error limits) compatible
with the [O~I] result.  

\section{Correlations with other stellar parameters}
The scatter of the deprojected flow velocities of the
forbidden lines is clearly greater than can be explained by the observational
errors alone. Thus, assuming that the disk inclination errors cited in the 
literature are correct, we have to conclude that significant,
real differences exist between the outflow velocities
of the individual objects. To clarify the origin of these
differences, we searched 
for correlations between the individual deprojected 
velocities and other stellar
properties. In detail, we checked for possible correlations between these
velocities and the objects' escape velocities, photospheric radii, 
rotation periods, magnetic field data, spectral veiling, mass accretion rates,
wind mass loss rates, and various line profile properties. No 
convincing correlations could be found. A trend, shown in Fig.~14, was 
found for the
relation between the deprojected [O I] velocity and the wind mass loss rate
as listed by Hartigan et al. (\cite{hartigan95}). The flow velocity appears to
increase systematically from about 200 km/s 
at $\dot{M} = -10^{-10} $M$_\odot$/yr to
$\approx $ 400 km/s at about $\dot{M} = -3 \times 10^{-7} $M$_\odot$/yr. 
However, this result is only based on eight data points. Perhaps more
remarkable may be that the velocity 
change is only a factor two, while the mass loss rate changes by 
about $3 \times 10^{3}$. 

\begin{figure}
\label{F14}
\centering
\includegraphics[angle=0,width=10cm]{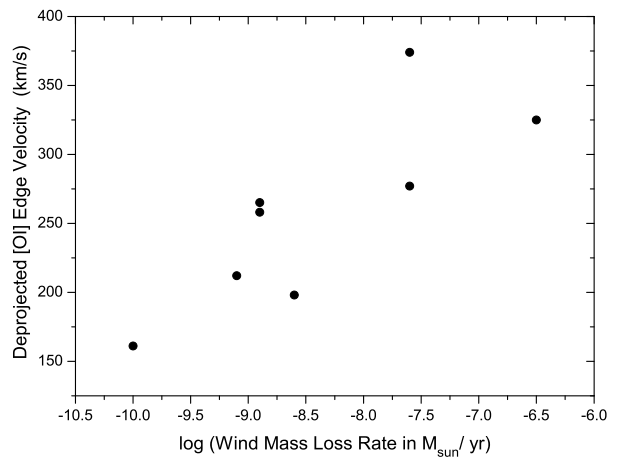}
\caption{Deprojected wind flow velocities inferred from the blue edge 
of the [OI] 630 nm lines as a function of the logarithm of the 
wind mass loss rate.}
\end{figure}

That no other correlations of the
deprojected velocity with stellar parameters could be detected does not
rule out that such correlations exist. In most cases the 
detection of significant correlations was precluded by the smallness of our
samples and/or by too large errors of the input data. Additional data will be needed  to find correlations with other stellar parameters and 
to clarify the origin of the differences in outflow velocities. 

\section{Conclusions}
The results described in the previous sections show that various 
properties of the spectra of CTTSs are correlated
with the inclinations of their associated circumstellar disks.
No such correlations could be detected for inclinations
derived from the rotational data of these stars, and the 
correlation between the rotation-based inclinations and the
disk inclinations is rather poor. This seems to indicate that only
the disk inclinations provide a good measure of the true orientations
of the TTS systems, while the values derived from stellar 
rotation data are not yet accurate enough for such studies.
      
The most conspicuous correlations with the disk inclination are observed 
for the profiles of the forbidden emission lines and for the 
velocities of the wind absorption features 
of some permitted lines. The data support the assumption of cone-like
fast polar winds with opening angles $\leq 45^\circ$. 
A correlation is also indicated between the disk inclination and the
width of the He~I 587.6 nm emission line. This supports earlier suggestions
that this line is at least in part related to the T~Tauri winds. 

Using the disk inclination values, true (deprojected) wind flow 
velocities were calculated. From the blue edge of the [O~I] emission,
we find mean flow velocities of about 260~km/s, with the 
well determined individual values ranging between 161~km/s and  
374~km/s. The blue edges of the wind absorption features of 
the He~I~1083~nm line give similar mean values, although the scatter 
of the individual velocities is greater. The deprojected central 
velocities of the H$\alpha$ wind absorption components have velocities
that are similar to or slightly larger than those derived from 
the blueshifted emission peaks of the [N~II] lines. The derived 
deprojected flow velocities differ among the observed objects.
Correlations of these differences with other stellar parameters could 
not be reliably established with the samples that are presently 
available. Larger samples, which will probably become available 
with future ALMA observations of TTS disks, may provide answers  
to this question.           

\begin{acknowledgements}
We would like to thank the referee, Dr. Suzan Edwards, for valuable 
comments and helpful suggestions. 
\end{acknowledgements}


\begin{thebibliography}{}

   \bibitem[2011]{akeson11} Akeson, R. L., et al. 2011,
      ApJ, 728, 96

   \bibitem[2000]{alencar00} Alencar, S. H. P. \& Basri, G. 2000,
      AJ, 119, 1881

   \bibitem[2002]{alencar02} Alencar, S. H. P. \& Batalha, C. 2002,
      ApJ, 571, 378

   \bibitem[2001]{alencar01} Alencar, S. H. P., Johns-Krull, C. M.,
      \& Basri, G. 2001,
      AJ, 122, 3335

   \bibitem[2007]{andrews07} Andrews, S. M. \& Williams, J. P. 2007
      ApJ, 659, 705

  \bibitem[2005]{appenzeller05} Appenzeller, I., Bertout, C. \&
      Stahl, O. 2005,
      A\&A, 434, 1005

  \bibitem[1986]{appenzeller86} Appenzeller, I., Jankovics, I. \&
      Jetter, R. 1986,
      A\&AS, 64, 65

   \bibitem[1989]{appenzeller89} Appenzeller, I., \& Wagner, S. 1989,
      A\&A, 225, 432

   \bibitem[2007]{azevedo07} Azevedo, R., et al. 2007,
      ApJ, 670, 1234

   \bibitem[2001]{beristain01} Beristain, G., Edwards, S., Kwan, J. 2001,
      ApJ, 551, 1037

   \bibitem[1988]{bertout88} Bertout, C., Basri, G., Bouvier, J. 1988,
      ApJ, 330, 350

   \bibitem[2007]{bertout07} Bertout, C., Siess, L., Cabrit, S. 2007,
      A\&A, 473, L21

  \bibitem[1984]{boesgaard84} Boesgaard, A. M. 1984,
      AJ, 89, 1635

   \bibitem[1994]{boehm94} B\"ohm, K.-H.. \& Solf, J. 1994,
      ApJ, 430, 277

   \bibitem[2007]{bouvier07} Bouvier, J., et al. 2007,
      A\&A, 463, 1017

   \bibitem[1995]{bouvier95} Bouvier, J., et al. 1995,
      A\&A, 299, 89

    \bibitem[2007]{broeg07} Broeg, C., et al. 2007,
      A\&A, 468, 1039

    \bibitem[1990]{camenzind90} Camenzind, M. 1990,
      Rev. Modern Astronomy, 3, 234

   \bibitem[2013]{cox13} Cox, A. W., et al. 2013,
      ApJ, 762, 40 

   \bibitem[1997]{curiel97} Curiel, S., et al. 1997,
      AJ, 114, 2736

   \bibitem[2003]{dutrey03} Dutrey, A., Guilloteau, S., Simon, M. 2003,
      A\&A, 402, 1003

   \bibitem[2012]{donati12} Donati, J.-F., et al. 20012,
      MNRAS, 425, 2948

   \bibitem[1994]{edwards94} Edwards, S., et al. 1994,
      AJ, 108, 1056

   \bibitem[2006]{edwards06} Edwards, S., et al. 2006,
      ApJ, 646, 319

   \bibitem[2008]{fischer08} Fischer, W. et al. 2008,
      ApJ, 687, 1117

   \bibitem[2012]{guilloteau12} Guilloteau, S., et al. 2012,
      A\&A, 529, 105

   \bibitem[1994]{hamann94} Hamann, F. 1994,
      ApJS, 93, 485

  \bibitem[1992]{hamann92} Hamann, F., \& Persson, S. E. 1992,
      ApJS, 82, 247

   \bibitem[2003]{hamilton03} Hamilton, C. M., et al. 2003,
      ApJ, 591, L45

   \bibitem[1995]{hartigan95} Hartigan, P., Edwards, S., 
     Ghandour, L. 1995,
     ApJ, 452, 736

   \bibitem[1994]{hartmann94} Hartmann, L., Hewett, R., \& Calvet, N. 1994,
      ApJ, 426, 669

   \bibitem[2007]{herczeg07} Herczeg, G. J., et al. 2007,
      ApJ, 670, 500

  \bibitem[1997]{hirth97} Hirth, G. H., Mundt, R., Solf, J. 1997,
      A\&AS, 126, 437

   \bibitem[2010]{isella10} Isella, A., et al. 2010,
      ApJ, 714, 1746

   \bibitem[2002]{jkrull02} Johns-Krull, C. M. \& Gafford, A. D. 2002,
      ApJ, 573, 685

   \bibitem[2013]{jkrull13} Johns-Krull, C. M., et al. 2013,
      ApJ, 765, 11

   \bibitem[1990]{krautter90} Krautter, J., Appenzeller, I.,
      Jankovics, I. 1990,
      A\&A, 236, 416

   \bibitem[2006]{kurosawa06} Kurosawa, R., Harries, T. J., \&
      Symington, N. H. 2006,
      MNRAS, 370, 580

   \bibitem[2011]{kurosawa11} Kurosawa, R., Romanova, M. M., \$ 
      Harries, T. J., 2011,
      MNRAS, 416, 2623

   \bibitem[2007]{kwan07} Kwan, J., Edwards, S., \& Fischer, W. 2007,
      ApJ, 657, 897

   \bibitem[2011]{kwan11} Kwan, J. \& Fischer, W. 2011,
      MNRAS, 411, 2383

   \bibitem[2011]{kwon11} Kwon, W., Looney, L. W. Mundy, L. G. 2011,
      ApJ 741, 3
   
   \bibitem[2001]{muzerolle01} Muzerolle, J., Calvet, N., \&
      Hartmann, L. 2001,
      ApJ, 550, 944

   \bibitem[2005]{osullivan05} O'Sullivan, M., et al. 2005,
      MNRAS, 358, 632

   \bibitem[2009]{ratzka09} Ratzka, T., et al. 2009,
      A\&A, 502, 623

   \bibitem[2011]{rocca11} Roccatagliata, V., et al. 2011,
      A\&A, 534, A33

   \bibitem[2009]{schegerer09} Schegerer, A. A., et al. 2009,
      A\&A, 502, 367

   \bibitem[1998]{shevchenko98} Shevchenko, et al. 1998,
      Astr.L. 24, 528

   \bibitem[1998]{stapelfeldt98} Stapelfeldt, K. R., et al. 1998,
      ApJ, 502, L65

   \bibitem[2003]{stapelfeldt03} Stapelfeldt, K. R., et al. 2003,
      ApJ, 589, 410

  \bibitem[2002]{stempels02} Stempels, H. C. \& Piskunov, N. 2002,
      A\&A, 391, 595

   \bibitem[2012]{tanii12} Tanii, R., et al. 2012,
      PASJ, 64, 124

   \bibitem[1999]{webb99} Webb, R. A., et al. 1999,
      ApJ, 512, L63
  
   \bibitem[1987]{weaver87} Weaver, W. B. 1987,
      ApJ, 319, L89

   \bibitem[1993]{yorke93} Yorke, H. W., Bodenheimer, B., Laughlin, G. 1993,
      ApJ, 411, 274


\end{thebibliography}
\end{document}